\documentclass[10pt,aip,showpacs,superscriptaddress, twocolumn, nofootinbib]{revtex4-1}
\usepackage{url}
\usepackage{amsmath,amssymb}
\usepackage{fourier}

\usepackage{amsmath,amssymb}
\usepackage[dvipdfm]{graphicx}
\usepackage{color}
\usepackage{setspace}

\newcommand{\vect}[1]{\mbox{\boldmath $#1$}}

\usepackage{url,ulem}

\begin{document}

\title{Study on creation and destruction of transport barriers via effective
safety factors for energetic particles}

\author{Shun Ogawa}
\email{shun.ogawa@cpt.univ-mrs.fr}


\affiliation{Aix Marseille Univ, Universit\'e de Toulon, CNRS, CPT, Marseille, France}

\affiliation{CEA, IRFM, F-13108 St. Paul-lez-Durance cedex, France }

\author{Xavier Leoncini}

\affiliation{Aix Marseille Univ, Universit\'e de Toulon, CNRS, CPT, Marseille, France}

\author{Guilhem Dif-Pradalier}

\affiliation{CEA, IRFM, F-13108 St. Paul-lez-Durance cedex, France }

\author{Xavier Garbet}

\affiliation{CEA, IRFM, F-13108 St. Paul-lez-Durance cedex, France }
\begin{abstract}
Charged particles with low kinetic energy move along magnetic field lines, but so do not energetic particles. 
We investigate the topological structure changes in the phase space of energetic particles with respect to the magnetic one. 
For this study cylindrical magnetic fields with non-monotonic safety factors that induce the magnetic internal transport barrier are considered. 
We show that the topological structure of the magnetic field line and of the particle trajectories can be quite different. 
We explain this difference using the concept of effective particle $q$-profile. 
Using this notion we can investigate the location and existence of resonances for particle orbits that are different from the magnetic ones. 
These are examined both numerically by integrating an equation of motion and theoretically 
by use of Alfv\'en's guiding center theory and by use of the effective reduced Hamiltonian for the integrable unperturbed system. 
It is clarified that, for the energetic particles, the grad $B$ drift effect shifts the resonances and 
the drift induced by curvature of the magnetic field line leads to the vanishing of the resonances. 
As a result, we give two different mechanisms that lead to the creation of transport barriers for energetic particles
in the region where the magnetic field line is chaotic.
\end{abstract}
\maketitle

\section{Introduction}

Understanding the motion of a charged particle motion in a complex magnetic field 
is one of the key ingredients of magnetically confined fusion plasmas \cite{Boozer2004} and it has been a long standing issue in plasmas physics. \cite{Alfven1940,Northrop1961,Littlejohn1981, Cary2009}
In this paper we focus our attention on the case of so called chaotic (also called stochastic) magnetic field lines, 
which typically occurs when due to a perturbation a magnetic surface breaks down. \cite{Rosenbluth}
Indeed if the perturbation consists of several modes, such that the created islands are overlapping, 
a global chaotic (stochastic) region can be expected in the magnetic field.\cite{Chirikov1979,Rechester}
Our study follows our recent study of full particle motion in magnetic fields that can give rise to so-called internal transport barriers (ITB).\cite{Ogawa2016} 
An ITB can be created between two chaotic regions 
by a non-monotonic\cite{Balescu1998} or a plateau\cite{Constantinescu2012} profiles of the safety factor $q_{{\rm mag}}(r)$, 
which induce reversed or low magnetic shears respectively.
In this perspective, the ITB emerges as a region foliated by invariant magnetic tori and acts as an effective barrier for particle transport that confines particles
when we assume that particles move along magnetic field lines. 
However, there exist several cases for which the topological structure of particle orbits is completely different from the magnetic one. 
These structure can change as well drastically by simply modifying the initial pitch angle or energy.\cite{Ogawa2016}
In fact it has been shown that when the energy is sufficiently large the effective location of the resonance
which was destroying the magnetic structure can be displaced and even disappear as we shall discuss later on.\cite{Fiksel2005,Gobbin2008, White2010}
To be more specific, when we consider only particles with small energy whose motion is mostly one along a magnetic field line and some gyration around it. 
Then, the qualitative structure (\textit{e.g.} place of the resonance point on which the winding number of orbit is rational)
is similar to the one of magnetic field lines, so that the magnetic ITBs \cite{Balescu1998, Constantinescu2012} act as the barrier for particles.
Meanwhile, for energetic particles, 
it was found by computing full particle orbits numerically that these structures are quite different from the magnetic ones.
\cite{Weitzner1999,Cambon2014,Ogawa2016} 
It is for instance possible that for a given class of particles with a range of energies and pitch angles, 
an ITB appears in the chaotic region of the magnetic field lines.\cite{Ogawa2016} 
Conversely it is also possible that the magnetic ITB is essentially useless from
the particles point of view.\cite{Ogawa2016} 

In this article, we focus on the creation of an effective ITB in the particle orbits, 
and we try to monitor them by checking the resonance shift and its potential disappearance. 
At the same time, we aim to shed some light on the numerical results discussed in Ref. \onlinecite{Ogawa2016}, 
using a more systematic theoretically predictive approach, which in some cases is not only qualitative, but also quantitative. 

This article is organized as follows. The model for self-consistency
a short recollection of the results obtained in Ref.~\onlinecite{Ogawa2016}
are presented in Sec.~\ref{sec:model}. In Sec.~\ref{sec:safety},
we discuss the notion of effective safety factor that we can compute
from particle trajectories. This effective $q$-profile allows us
to reason using a classical intuition that particle follow field lines,
in that sense it appeared as a powerful tool. This effective profile
is computed numerically, but also theoretical expressions are obtained,
one from the full particle orbits in the integrable case, and
the other from the guiding center orbits approximation. 
We then analyze the consequence of these computations and the resonance shift is numerically checked in Sec.~\ref{sec:num}. 
We can then conclude on the influence of energy and pitch angle on the effective $q$-profile and understand 
or predict which effect induces a resonance shift or its disappearance.
Finally we conclude in Sect.~\ref{sec:Conclusion}.

\section{Model and basic setup\label{sec:model}}

The setting we consider is the one of a cylindrical set up, meaning
that we assume that we have magnetic field lines that wind around
concentric cylinders and the winding is depending only on the radial
coordinate $r$. Also and for simplicity we assume that our cylinder
is as well $2\pi R_{{\rm per}}$ periodic along its in $z-$axis. 
The natural choice of coordinates is then the cylindrical ones $(r,\theta,z)$.
Due to the divergent free nature of the magnetic field, 
field lines can be described using a Hamiltonian formalism, and in this original setting, 
the Hamiltonian is integrable. 
Then if we consider a magnetic perturbation corresponding to a specific Fourier mode $e^{i(m\theta-nz/R_{{\rm per}})}$
(due to the periodicity of the model), this mode, if isolated, creates what is called a magnetic island located around the resonant
cylinder at radius $r$ for which the safety factor (winding number) is such that $q_{{\rm mag}}(r)=m/n$. 
We may add more modes in the perturbation to create Hamiltonian chaos in magnetic field lines. 

In order to perform some computations, we simplify our notation using dimensionless variables, 
in this setting the $2\pi R_{{\rm per}}$-periodic boundary condition in the axis of cylinder, 
and we can set without loss of generality $R_{{\rm per}}=1$ 
when we for instance choose a radius of the cylinder as our units $r_{{\rm cyl}}=1$. 
We denote $e$ and $M$ the charge and mass of a particle respectively, 
and $B_{0}$ is the typical value of the magnetic filed on the axis. 
The conserved kinetic energy of the particle is $E=m\|\vect{v}\|^{2}/2$ where $\vect{v}$ 
denotes a velocity of the particle and $\|\bullet\|$ denotes Euclidian norm in $\mathbb{R}^{3}$. 
As mentioned earlier, to simplify the notations, 
we rescale the space $\vect{x}$ and time $t$ as $\tilde{\vect{x}}=\vect{x}/r_{{\rm cyl}}$, $\tilde{t}=t/\omega_{{\rm gyr}}$ respectively, 
where the gyro-frequency $\omega_{{\rm gyr}}=|e|B_{0}/M$, as done in Refs. \onlinecite{Cambon2014, Ogawa2016}.
The rescaled velocity is $\tilde{\vect{v}}={\rm d}\tilde{\vect{x}}/{\rm d}\tilde{t}$, and the rescaled energy $\tilde{E}=\|\tilde{\vect{v}}\|^{2}/2$ 
which relates to the particle energy $E$ as 
\begin{equation}
	E=\frac{|e|^{2}B_{0}^{2}r_{{\rm cyl}}^{2}}{M}\tilde{E}.
\end{equation}
Thus we can estimate the value of the energy in keV for the alpha particle or proton 
by a multiplication by $10^{5}$ of the rescaled value $\tilde{E}$ when $B_{0}=1\,T$ and $r_{{\rm cyl}}=1\,m$. 
In the rest of the paper, all values of energy are given in as values of $\tilde{E}$, and we omit tildes. 
We now consider the motion of a charged particle in this cylindrical magnetic field $\vect{B}(r,\theta,z)$
associated with the Coulomb gauge vector potential 
\begin{equation}
	\begin{split}
		\vect{A}(r,\theta,z) & =\vect{A}_{0}(r)+\epsilon\vect{A}_{1}(r,\theta,z),\\
		\vect{A}_{0}(r) & =-B_0\left( \frac{r}{2}\vect{e}_{\theta}+F(r)\vect{e}_{z}\right),\\
		\epsilon\vect{A}_{1}(r,\theta,z) & =\epsilon\vect{e}_{z}\sum_{m,n}A_{1}^{m,n}(r)e^{i(m\theta-nz/R_{{\rm per}})},
	\end{split}
\label{eq:model}
\end{equation}
where $\vect{e}_{i}$ are the basic unit vectors for each direction,
$i=\theta,z$, and where $F(r)$ is given by 
\begin{equation}
	F(r)=\int^{r}f(r){\rm d}r,\quad f(r)=\frac{r}{R_{{\rm per}}q_{{\rm mag}}(r)}.
\end{equation}
This motion is governed by the Hamiltonian 
\begin{equation}
	H(\vect{q},\vect{p})=\frac{\|\vect{p}-\vect{A}(\mathbf{q})\|^{2}}{2}\:.
	\label{eq:Ham}
\end{equation}

\begin{figure}[tb]
	\centering{}\includegraphics[width=8cm]{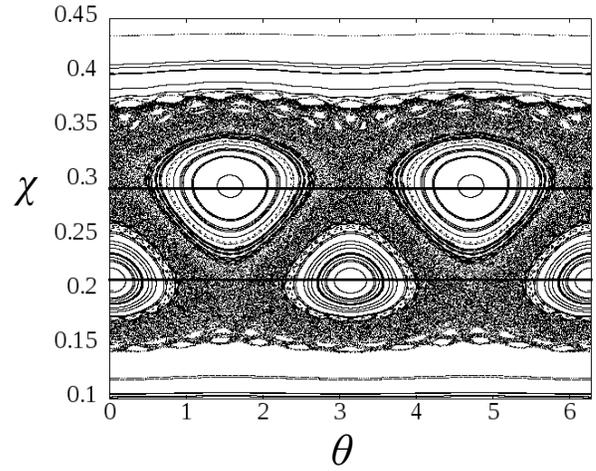} 
	\caption{
		Profile of the magnetic field line. 
		The Poincar|'e plot is taken for each $z\in2\pi R_{{\rm per}}\mathbb{N}$. 
		Two bold lines  $\chi=169/576,121/576$ correspond to the resonance 
		$q_{{\rm mag}}(r)=2/3$. \label{fig:magnetic} }
\end{figure}

\begin{figure}[bt]
	\centering{}
	(a)\\
 	\includegraphics[width=8cm]{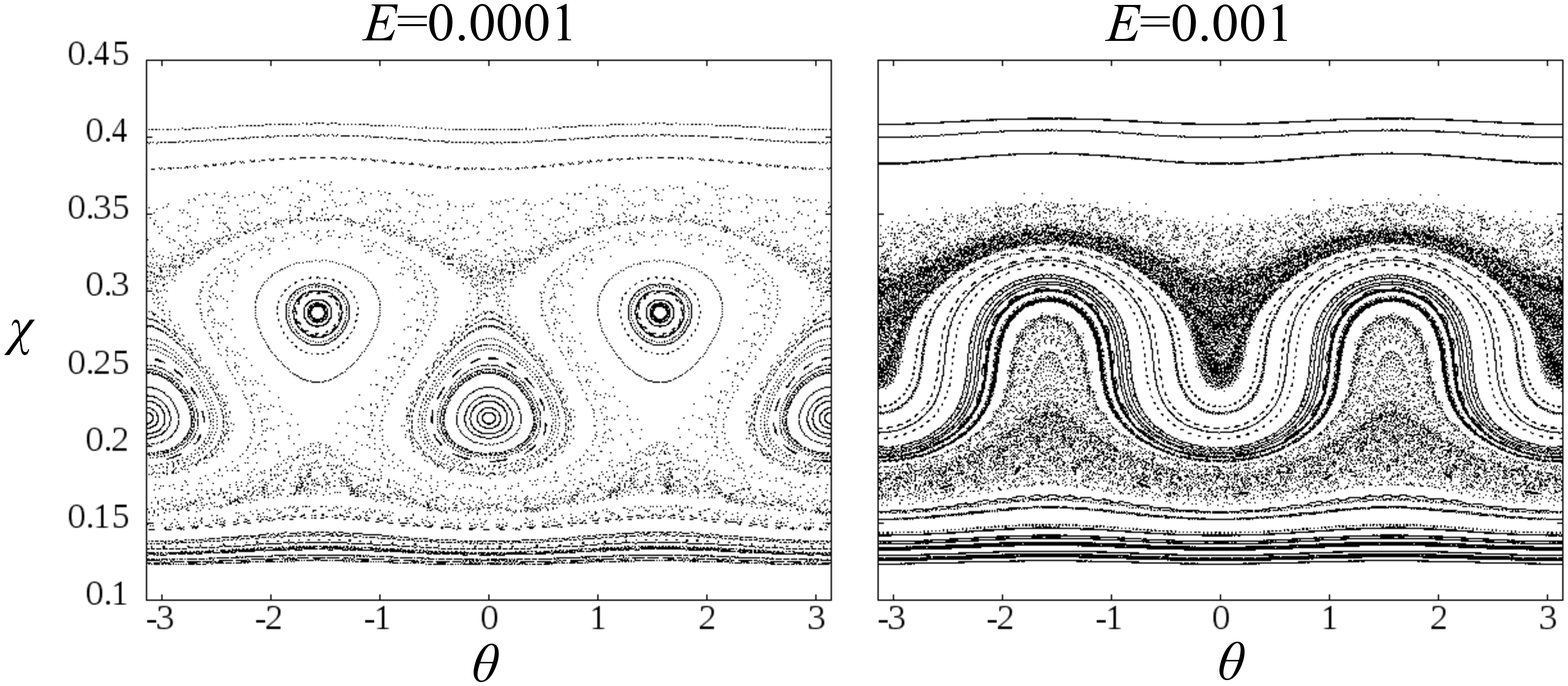}\\
 	(b)\\
 	\includegraphics[width=8cm]{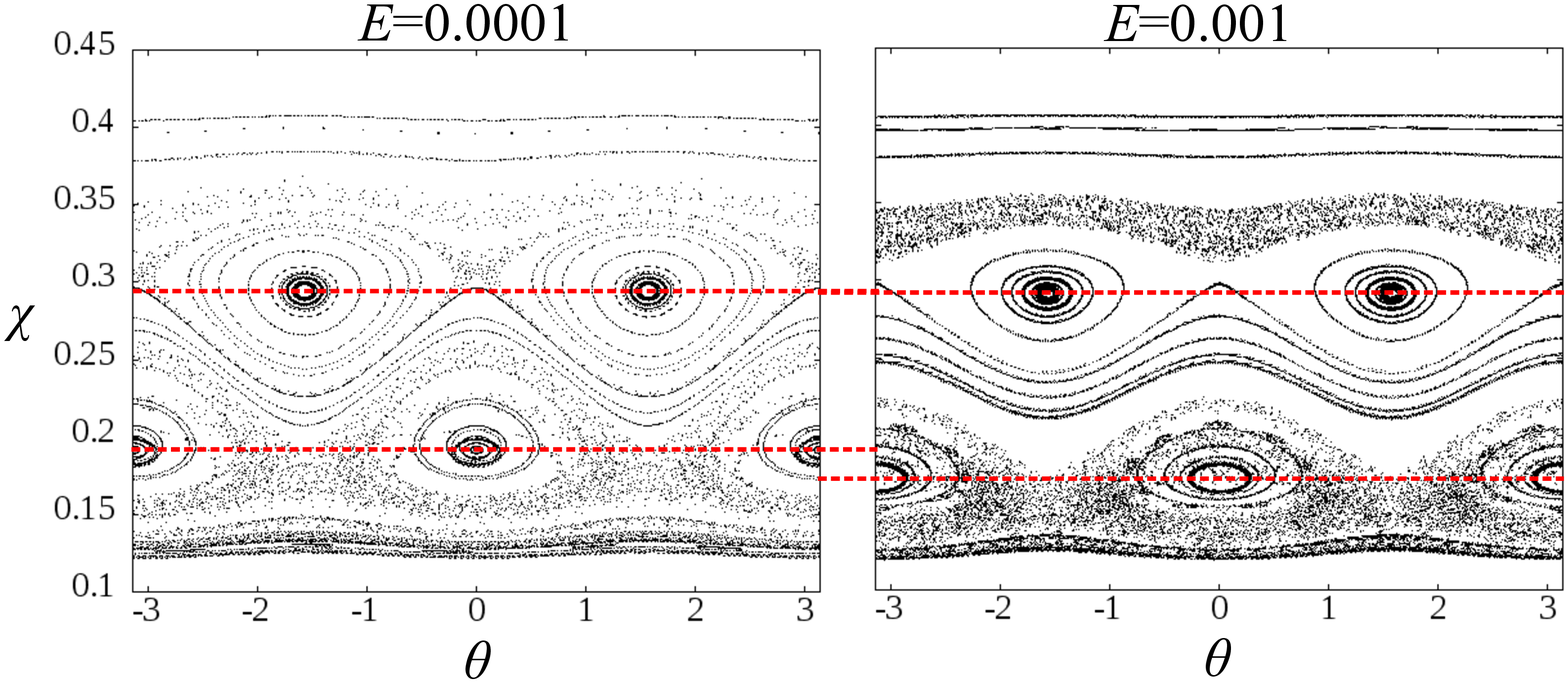} 
 	\caption{ (Color online) 
		The panels (a) and (b) are particle orbit profiles in the magnetic field lines exhibited 
		in Fig.~\ref{fig:magnetic} for null pitch angle and for pitch angle 1.25 rad respectively. 
		These snapshots are taken for the same plane with the magnetic field profile. 
		The panel (a) shows that the resonances corresponding to $q=2/3$ disappear when the energy increases. 
		In the panel (b), red broken lines indicate resonances.
		The resonances shift to the inside of the cylinder as energy increases.
	}
	\label{fig:particle-orbits} 
\end{figure}

Since we are interested in internal transport barrier (ITB), 
we consider a non-monotonic $q$-profile which creates magnetic ITB.\cite{Balescu1998}
To be more specific, we picked a safety factor $q_{{\rm mag}}(r)$ which is used in the study,\cite{Ogawa2016}
\begin{equation}
	q_{{\rm mag}}(r)=q_{0}\left[1+\lambda^{2}\left(r-\alpha\right)^{2}\right],\label{eq:safety}
\end{equation}
where $q_{0},\alpha$, and $\lambda$ are some constants. 
In this article we set $q_{0}=0.64$, $\alpha=1/\sqrt{2}$, and $\lambda=3$.
Because of the non-monotonicity of $q_{{\rm mag}}$, there are two resonance magnetic surfaces for a given rational $q_{{\rm mag}}$.
To clarify some ideas we exhibit a Poincar\'e plot of the magnetic field lines and the several particle orbits profiles. 
We consider the perturbation\cite{Ogawa2016}
\begin{equation}
	\begin{split} 
		& \epsilon A^{2,3}(r)\cos(2\theta-3z/R_{{\rm per}})\\
		& +\epsilon cA^{13,17}(r)\cos(13\theta-17z/R_{{\rm per}}),
	\end{split}
\end{equation}
where $\epsilon=0.0015$ and $c=0.02$. 
The amplitude $A^{m,n}(r)$ of the mode has peaks at $r$ satisfying $q_{{\rm mag}}(r)=m/n$, and is squeezed around there. 
The Poincar\'e plot of the magnetic field line on $(\theta,\chi)$ plane is exhibited in Fig.~\ref{fig:magnetic},
where the flux $\chi=B_{0}r^{2}/2$. 
And, several particle orbit profiles are shown in Fig.~\ref{fig:particle-orbits}, 
there and one can observe the resonance shift for large pitch angles and its disappearance for small pitch angles.

Before moving on we summarize the numerical observations made in Ref.~\onlinecite{Ogawa2016} in the Table \ref{tab:sum}. 
\begin{table}[b]
    \centering
	\renewcommand{\arraystretch}{1.25}
	\caption{
    		Summary of topology change: 
		Phenomena (2) and (3) have been already 
		explained theoretically. 
		The phenomenon (1) has not been clarified yet. 
    	}

	\begin{tabular}{r||c|c}
		\hline
		\hline
			& Small initial pitch angle & Large initial pitch angle  \\
		\hline
		\hline
		\shortstack{High\\energy}& 
		\multicolumn{2}{|c}{
			\shortstack{
			(1) Effect of magnetic field perturbation is 
			}
		}
		\\ \cline{1-2}
		\shortstack{$\qquad$\\ {\huge $\Uparrow$}\\$\qquad$}& 
		\shortstack{
			(2) Effect of magnetic field \\ perturbation 
			in trajectory \\
			becomes larger.
		}
		&
		\shortstack{suppressed, and \\
			ITB appears and \\ gets to be wider.
		}
		\\
		\hline
		\shortstack{Low\\energy}&  
		\multicolumn{2}{|c}{(3) Particles move along magnetic field line.}
		\\
		\hline
		\hline
	\end{tabular}
	     \label{tab:sum}
	\caption{
		Summary of topology change: 
		Phenomena (2) and (3) have been theoretically discussed in Ref.~\onlinecite{Ogawa2016}. 
		The phenomenon (1) is not yet clear.
	}
\end{table}
In fact the perturbation effects are enhanced as energy is increased for initial condition with a small pitch angle 
and not so energetic particle ($E \sim 10^{-4}$). 
This is because, in this case, 
the guiding center orbit of the particle in the unperturbed magnetic field is not so different from the field line 
as we will see in Sec.~\ref{sec:num}, 
and then the modification induced 
by the perturbation is dominant in the Lorentz force $\vect{F}_{{\rm L}}=\vect{v}\wedge\vect{B}$.\cite{Ogawa2016} 
We also were able to explain why the perturbation effects appears to get wiped out as energy increases when the pitch angle is large, 
considering the averaging induced by finite Larmor radius effect.\cite{Castillo2012,Martinell2013}

In this paper we shall reconsider these results from another perspective,
namely we shall consider some effective safety factor of a particle trajectory. 
The idea is heavily inspired from what was dubbed the ion guiding center safety factor in Ref.~\onlinecite{Fiksel2005}. 
As we shall see, using this point view, most of the behaviors observed in Ref.~\onlinecite{Ogawa2016}
can be understood.

\section{Effective Safety factors \label{sec:safety}}

In order to analyze the typical behavior of trajectories observed on a Poincar\'e section, 
we consider another typical measurement following what was suggested in Ref.~\onlinecite{Fiksel2005} 
and we consider the notion of an effective safety factor in our cylindrical torus (periodic cylinder) for a particle orbit given by 
\begin{equation}
	q_{{\rm eff}}\left(\langle r\rangle_{T}\right)=\frac{\langle\dot{z}\rangle_{T}}{R_{{\rm per}}\langle\dot{\theta}\rangle_{T}},\quad(R_{{\rm per}}=1)
\end{equation}
where $\dot{\,}$ denotes the time derivative, {\it i.e.} $\dot{a}={\rm d}a/{\rm d}t$,
and the brackets with subscript $T$ denote time average, $\langle a\rangle_{T}\equiv\frac{1}{T}\int_{0}^{T}a(t){\rm d}t$.
To shed some light on this quantity, we use can compute an approximation
of this value $q_{{\rm eff}}^{{\rm gc}}(r)$ obtained from Alfv\'en's guiding center velocity\cite{Boozer2004} 
which we write as $\vect{v}_{{\rm gc}}=v_{{\rm gc}}^{z}\vect{e}_{z}+r\Omega_{{\rm gc}}^\theta \vect{e}_{\theta}$.
In order to be more explicit, let us give some arguments on why the winding number of the particle is more relevant than the magnetic one.
For this purpose we do not need the details of the guiding center equation for $\epsilon>0$,
and using some abstract expression is enough. The evolution of the guiding center $\vect{x}_{{\rm gc}}$ is derived by solving the differential equation 
\begin{equation}
	\frac{{\rm d}\vect{x}_{{\rm gc}}}{{\rm d}t}=\vect{v}_{{\rm gc}}.
\end{equation}
We divide the vector field into the unperturbed part and the perturbation part as 
\begin{equation}
	\vect{v}_{{\rm gc}}=\vect{v}_{{\rm gc}}^{0}(r)+\epsilon\vect{v}_{{\rm gc}}^{1}(r,\theta,z).
\end{equation}
Recalling that the perturbation term of the field takes the form (\eqref{eq:model}), the perturbation term of the velocity field is written similarly as
\begin{equation}
	\epsilon\vect{v}_{{\rm gc}}^{1}(r,\theta,z)=
	\epsilon\sum_{m,n}\hat{\vect{v}}_{{\rm gc}}^{1,m,n}(r)e^{i(m\theta-nz/R_{{\rm per}})}.
\end{equation}
The solution to the unperturbed equation with initial condition $r_{0}(t=0)=\bar{r}_{0}$,
$\theta_{0}(0)=\bar{\theta}_{0}$, $z_{0}(t)=Z_{0}$, $\dot{\theta}_{0}(0)=\Omega_{{\rm gc}}^{\theta,0}(\bar{r}_{0})$
can be written as 
\begin{equation}
	\begin{split}
		r_{0}(t) & =\bar{r}_{0},\\
		\theta_{0}(t) & =\Omega_{{\rm gc}}^{\theta,0}(\bar{r}_{0})t+\bar{\theta}_{0},\\
		z_{0}(t) & =v_{{\rm gc}}^{z,0}(\bar{r}_{0})t+Z_{0}.
	\end{split}
\end{equation}
Thus, a secular term appears for $r$ satisfying 
\begin{equation}
	q_{{\rm eff}}^{{\rm gc}}(r)=\frac{v_{{\rm gc}}^{z,0}(r)}{R_{{\rm per}}\Omega_{{\rm gc}}^{\theta,0}(\bar{r}_{0})}=\frac{m}{n}
\end{equation}
for $m,n\in\mathbb{Z}$ with $\hat{\vect{v}}_{{\rm gc}}^{1,m,n}(r)\neq\vect{0}$.
As a consequence, in this approximation the resonance for the particle is not where $q_{{\rm mag}}(r)\in\mathbb{Q}$ but on the cylinder(s)
for which $q_{{\rm eff}}^{{\rm gc}}(r)\in\mathbb{Q}$. 
Given this phenomenon, we shall investigate further in order to see 
if this effect can explain the topological changes in particle trajectories observed in Ref.~\onlinecite{Ogawa2016}. 
For this purpose we start by pushing a bit further our computations and obtain some theoretical expressions of the effective $q$-profile.

\section{Theoretical expressions of the effective $q$-profile\label{sec:model-3}}

In this section we give an estimation of the difference between $q_{{\rm eff}}(r)$
and $q_{{\rm mag}}(r)$, two analytical expressions are obtained. 
One is computed by using the previously derived explanation and pursuing
the approach using the Alfv\'en's guiding center theory and the other
expression is computed using a the long-time average of full particle
orbits in the integrable approximation. We start with the latter.

\subsection{Full particle orbit}

We concern the long-time average of full particle orbit. 
The velocity elements $\dot{\theta}$ and $v_{z}=\dot{z}$ are can be guessed from the Hamiltonian \eqref{eq:Ham}
which corresponds to a rewriting of the kinetic energy; we thus have 
\begin{equation}
	r^2\dot{\theta}=P_{\theta}-\frac{B_{0}r^2}{2},\quad v_{z}=p_{z}+F(r).
\end{equation}
To go any further we assume that the system is integrable so we are
now considering the case when $\epsilon=0$, so that 
$P_\theta$ 
and $p_{z}$ are invariants of motion. Then , we have 
\begin{equation}
\langle\dot{\theta}\rangle_{\infty}=P_{\theta}\left\langle r^{-2}\right\rangle _{\infty}-\frac{B_{0}}{2},\quad\langle v_{z}\rangle_{\infty}=p_{z}+\langle F(r)\rangle_{\infty},\label{eq:timeave-v}
\end{equation}
where 
\begin{equation}
\begin{split}F(r)= & \frac{1}{R_{{\rm per}}}\int_{0}^{r}\frac{r'}{q(r')}{\rm d}r'\\
= & \frac{1}{R_{{\rm per}}q_{0}}\left[\frac{\ln\left(1+\lambda^{2}\left(r-\alpha\right)^{2}\right)}{2\lambda^{2}}+\frac{\alpha\arctan\left(\lambda(r-\alpha)\right)}{\lambda}\right].
\end{split}
\end{equation}
To determine the time average, we need an orbit associated with 
an effective Hamiltonian $H_{{\rm eff}}$ defined on $(r,p_{r})$ plane,\cite{Ogawa2016,Cambon2014}
\begin{equation}
\begin{split}
	H_{{\rm eff}}(r,p_{r}) & =p_{r}^{2}/2+V_{{\rm eff}}(r),\\
	V_{{\rm eff}}(r) & =\frac{P_{\theta}^{2}}{2r^{2}}+\frac{B_{0}^2r^{2}}{8}+\frac{(F(r)+p_{z})^{2}}{2}-\frac{B_{0}P_{\theta}}{2}.
\end{split}
\end{equation}
Since $H_{{\rm eff}}$ is one-degree of freedom, it is integrable,
and due to the shape of the effective potential all trajectories 
(except maybe for a few isolated fixed points) are periodic, 
so the long-time average can be replaced with an average over a period of the trajectory $T_{\rm gyr}$; 
for a given trajectory periodic trajectory with effective energy $E$ we can expect to get two extremal values for $r$ 
(unless we are on a circle) that are joined over half a period, then using ${\rm d}t={\rm d}r/p_{r}$
we just have to compute\cite{Gradov1995}  
\begin{equation}
	\begin{split}
		\langle g(r)\rangle_{\infty} & =\lim_{\tau\to\infty}\frac{1}{\tau}\int_{0}^{\tau}g\left(r(t)\right){\rm d}t\\
		 & =\frac{1}{T_{\rm gyr}}\int_{0}^{T}g\left(r(t)\right){\rm d}t
		  =\frac{2}{T_{\rm gyr}}\int_{r_{\rm min}}^{r_{\rm max}}\frac{g(r)}{p_{r}}dr\\
		 & =\frac{\int_{r_{\rm min}}^{r_{\rm max}}\frac{g(r)dr}{\sqrt{E-V_{\rm eff}(r)}}}{\int_{r_{\rm min}}^{r_{\rm max}}\frac{dr}{\sqrt{E-V_{\rm eff}(r)}}}\:.
	\end{split}
	\label{eq:ergod}
\end{equation}
where $P_{\theta}$ and $p_{z}$ are given by the initial condition,
and where turning points $r_{{\rm min}}$ and $r_{{\rm max}}$ are
such that $E-V_{{\rm eff}}(r_{{\rm min}/{\rm max}})=0$ and $r_{{\rm max}}>r_{{\rm min}}$.
We note that the one gyro-period is computed as
\begin{equation}
	\label{eq:Tgyr}
	T_{{\rm gyr}}=\sqrt{2}\int_{r_{{\rm min}}}^{r_{{\rm max}}}\frac{{\rm d}r}{\sqrt{E-V_{{\rm eff}}(r)}}.
\end{equation}
Substituting Eq. \eqref{eq:ergod} into Eq. \eqref{eq:timeave-v},
we get an analytical expression of the effective $q$-profile, 
\begin{equation}
q_{{\rm eff}}(\langle r\rangle_{\infty})=\langle v_{z}\rangle_{\infty}/(R_{{\rm per}}\langle\dot{\theta}\rangle_{\infty})
\end{equation}
To compute it, we need values of $P_{\theta}$ and $p_{z}$ which are given by initial velocity $\vect{v}_{0}$ and place $\vect{r}_{0}$ as 
\begin{equation}
p_{z}=v_{z}^{0}-F(r_{0}),\quad P_{\theta}=r_{0}^{2}\dot{\theta}_{\theta}^{0}+\frac{B_{0}r_{0}^{2}}{2},
\end{equation}
where $r_{0}=\|\vect{r}_{0}\|$, $v_{z}^{0}=\vect{v}_{0}\cdot\vect{e}_{z}$,
and $\dot{\theta}_{\theta}^{0}=\vect{v}_{0}\cdot\vect{e}_{\theta}/r_{0}$.
Then, we have $V_{{\rm eff}}$ and $r_{{\rm min/max}}$.

\begin{figure}[tb]
	\centering{}(a)\\
	 \includegraphics[width=8cm]{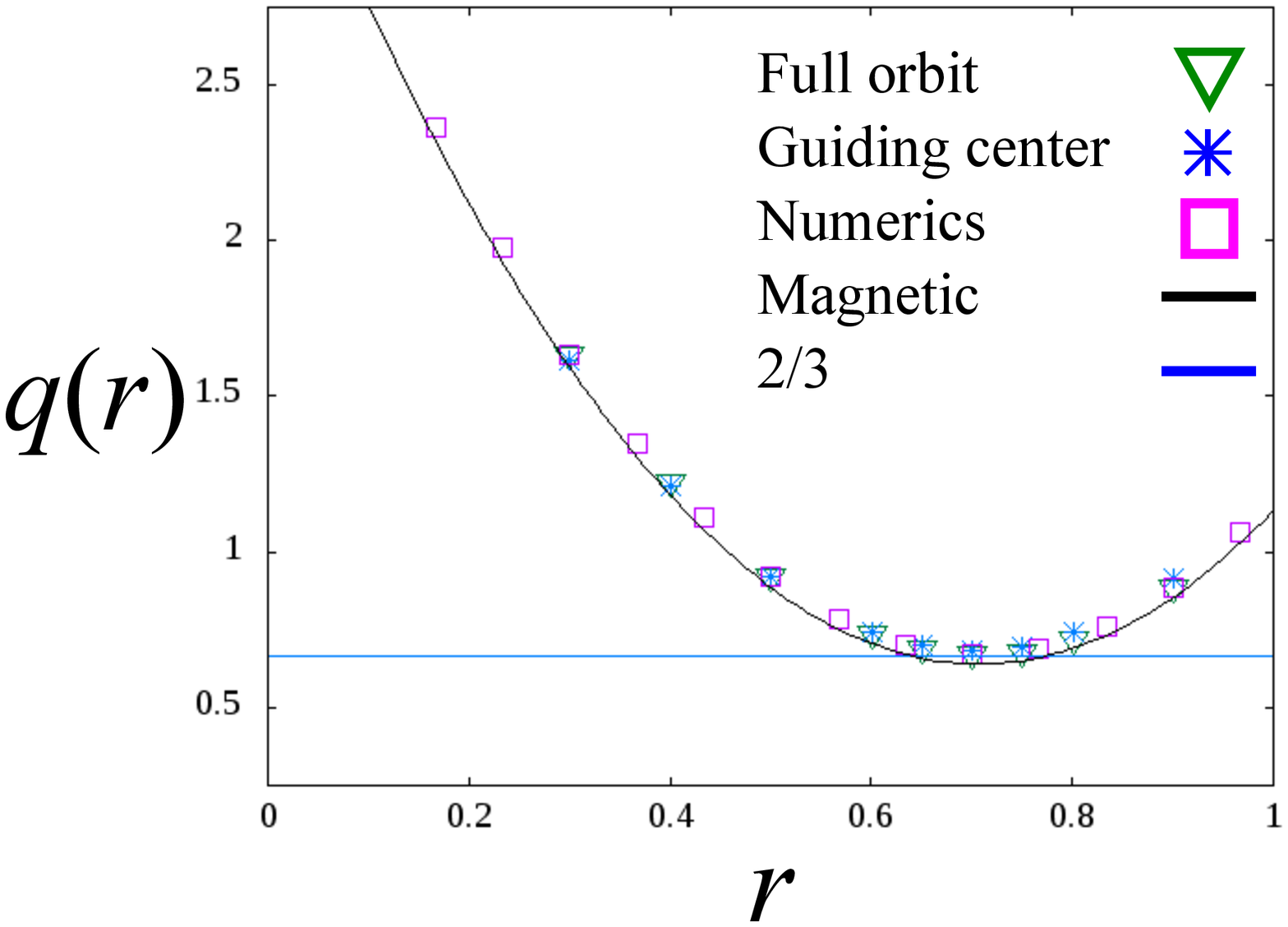}\\
	 (b)\\
	 \includegraphics[width=8cm]{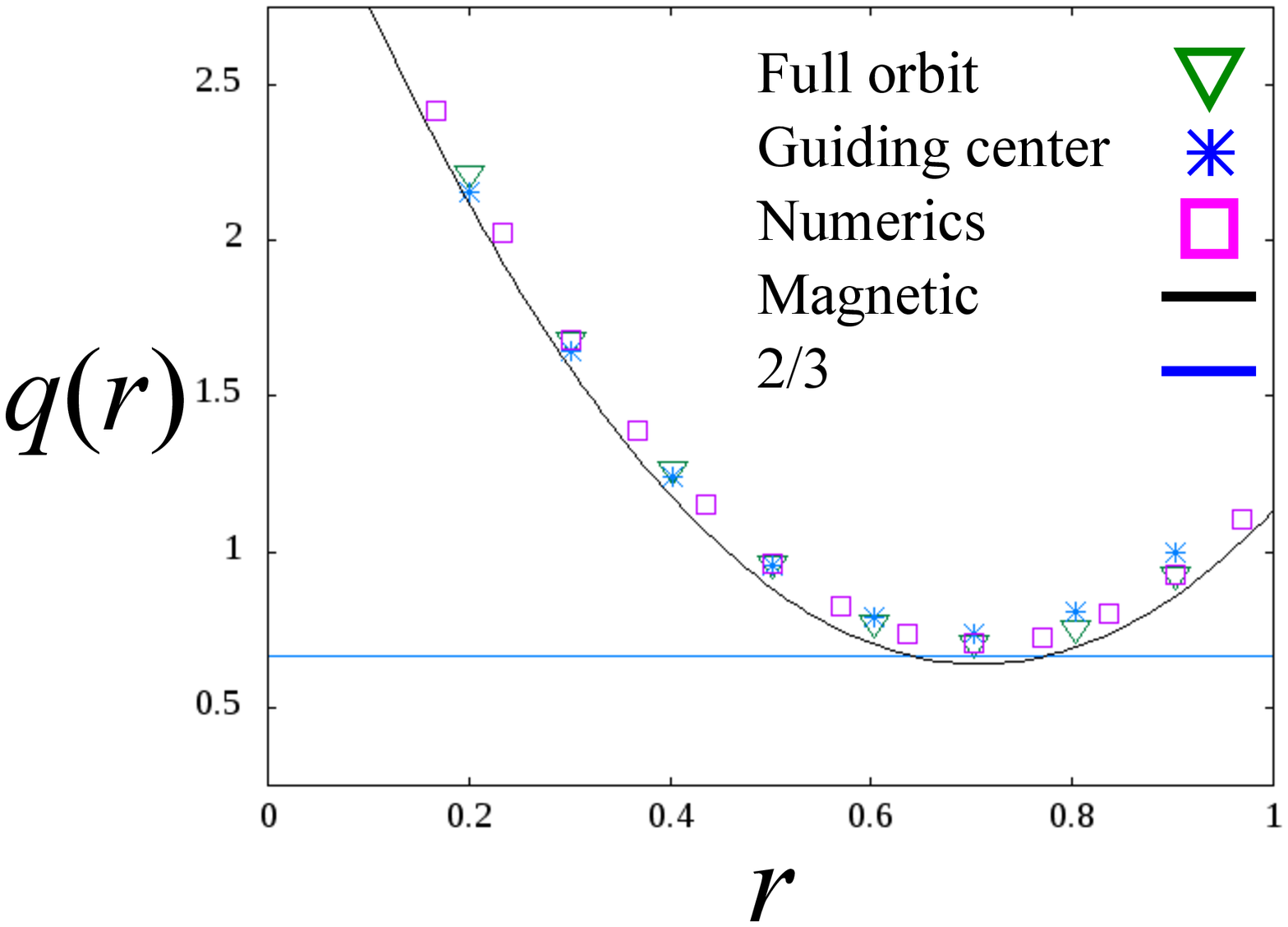} 
	 \caption{(Color online)
	 	$q_{{\rm mag}}$, $q_{{\rm eff}}(\langle r\rangle_{\infty})$, $q_{{\rm eff}}(\langle r\rangle_{T})$
		for finite $T$, and $q_{{\rm eff}}^{{\rm gc}}$ are plotted as functions of $r$. 
		The solid curve is $q_{{\rm mag}}$. The inverse triangles give $q_{{\rm eff}}(\langle r\rangle_{\infty})$ obtained theoretically,
		the stars $q_{{\rm eff}}^{{\rm gc}}$ obtained from the guiding center theory, 
		and the squares $q_{{\rm eff}}(\langle r\rangle_{T})$ which is obtained for finite $T$ numerically. 
		The energy are $E=0.001$ (top) and $E=0.005$ (bottom) respectively and initial pitch angle is $0$. 
		The blue straight line represents $q=2/3$.
	}
\label{fig:theory-null} 
\end{figure}

\begin{figure}[bt]
\centering{}
	(a)\\
	 \includegraphics[width=8cm]{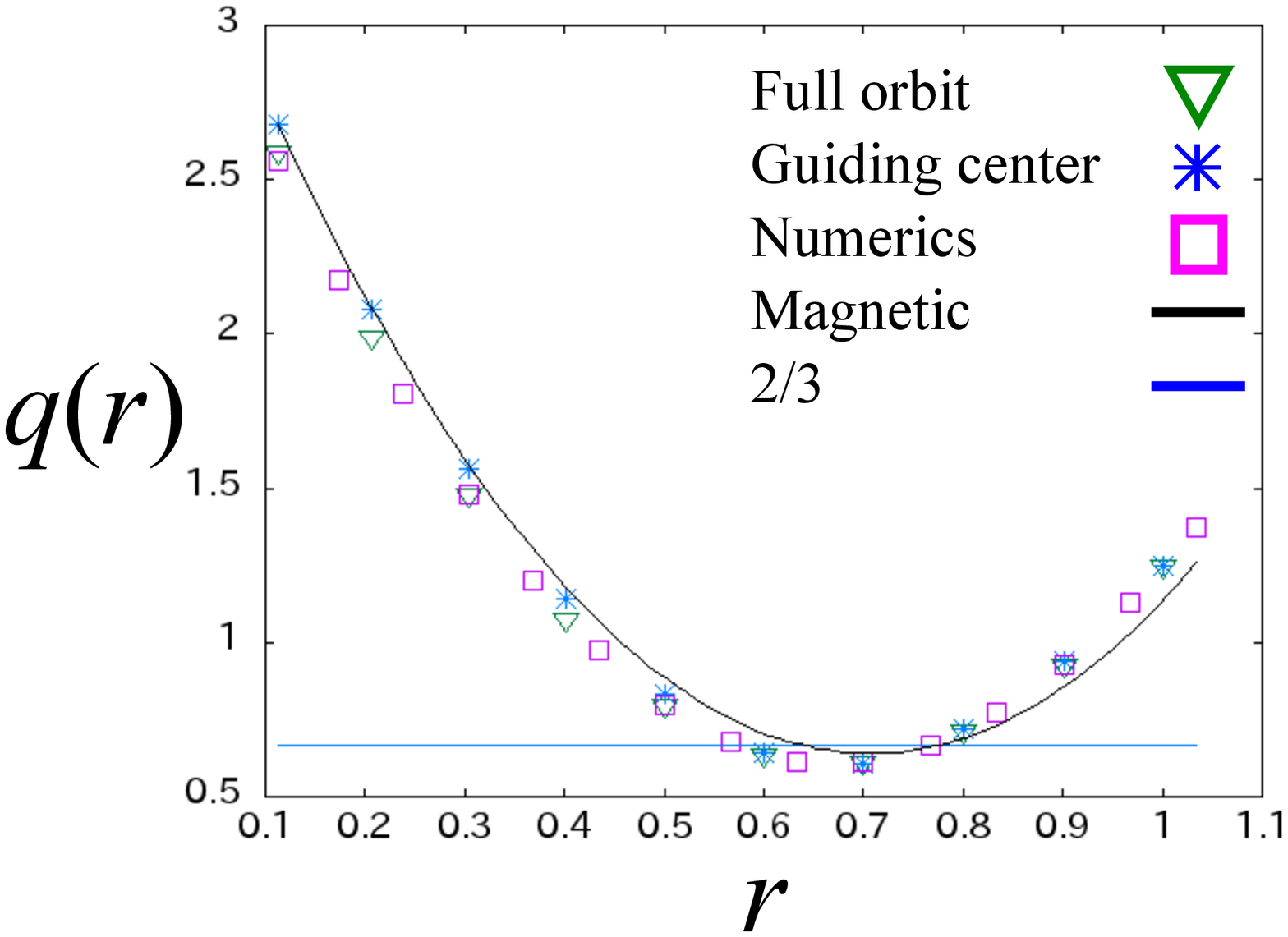}\\
	(b)\\
	\includegraphics[width=8cm]{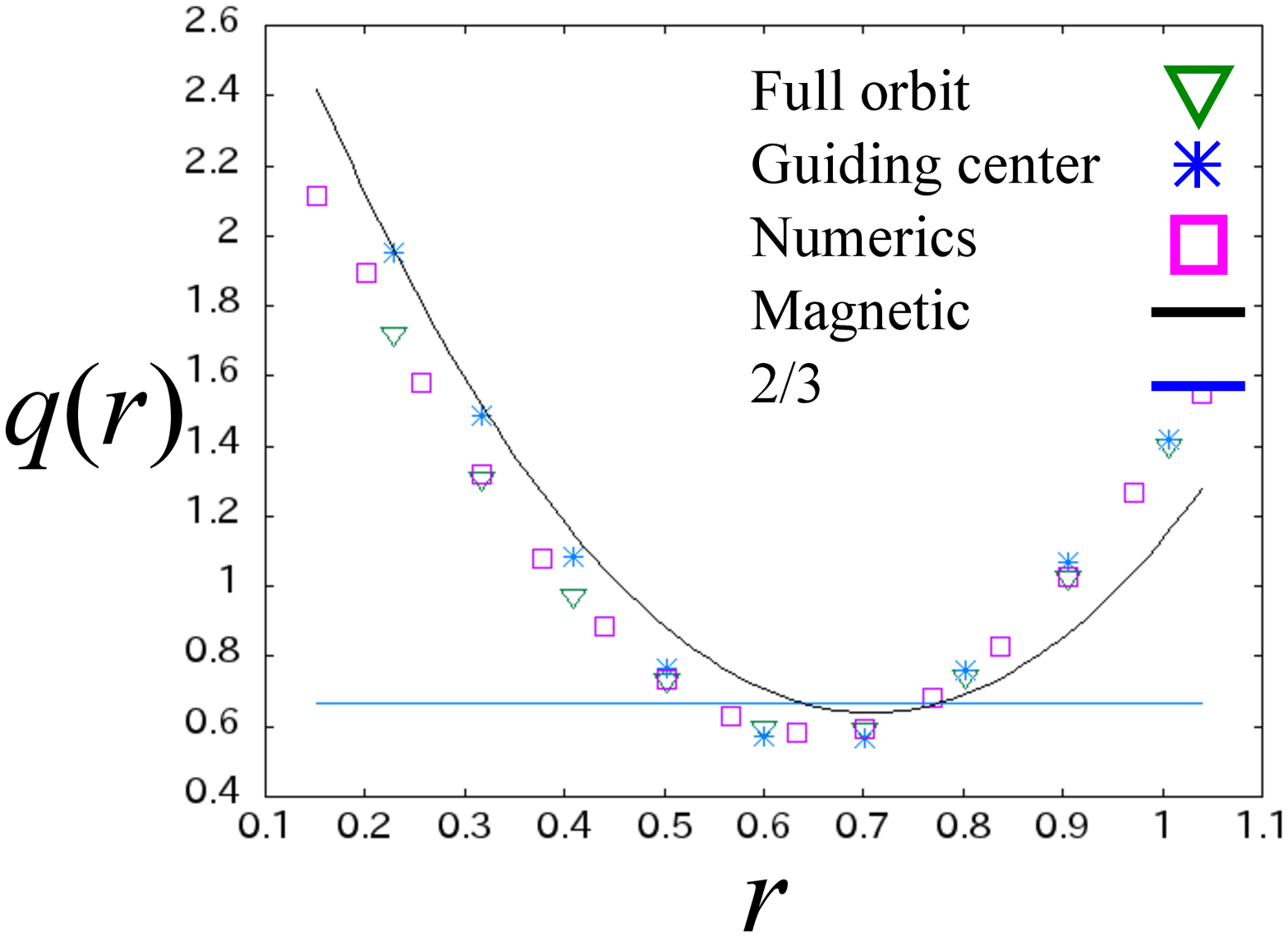} 
 	\caption{ 
		(Color online)
 		$q_{{\rm mag}}$, $q_{{\rm eff}}(\langle r\rangle_{\infty})$, $q_{{\rm eff}}(\langle r\rangle_{T})$ for finite $T$, 
		and $q_{{\rm eff}}^{{\rm gc}}$ are plotted as functions of $r$. 
		The solid curve is $q_{{\rm mag}}$. 
		The inverse triangles give $q_{{\rm eff}}(\langle r\rangle_{\infty})$ obtained theoretically,
		the stars $q_{{\rm eff}}^{{\rm gc}}$ obtained from the guiding center theory, 
		and the squares $q_{{\rm eff}}(\langle r\rangle_{T})$ which is obtained for finite $T$ numerically. 
		The energy are $E=0.001$ (top) and $E=0.005$ (bottom) respectively and initial pitch angle is $1.25$. 
		The blue straight line represents $q=2/3$. 
	}
\label{fig:theory-125e-2} 
\end{figure}

We compare the theoretically $q_{{\rm eff}}(r)$ and numerical ones
in Fig.~\ref{fig:theory-null} and \ref{fig:theory-125e-2}. 
The numerics for finite $T$ and the theory for $T\to\infty$ are in good agreement,
which in some sense confirms that our numerical algorithm using 
the symplectic sixth-order Gauss-Legendre scheme\cite{MacLchlan92} is correct.

\subsection{Guiding center calculation}

As another expression, we use Alfv\'en's guiding center approximation 
which is the oldest guiding center theory.\cite{Alfven1940, Boozer2004}
Although there exist progress after Alfven's seminal work\cite{Northrop1961, Littlejohn1981, Cary2009}, 
this is enough and convenient for our purpose. 
This allows us to compare with results already obtained from the integrable case. 
Using the formula\cite{Boozer2004} for a null electric field and a static magnetic field we get
\begin{equation}
	\vect{v}_{{\rm gc}}=v_{\parallel}\vect{b}+\frac{\vect{b}}{\omega_{{\rm gyr}}}\wedge\left(v_{\parallel}^{2}\vect{\kappa}+\frac{v_{\perp}^{2}}{2}\frac{\nabla B}{B}\right)
	\label{eq:vgc}
\end{equation}
where $B=\|\vect{B}\|$, $\vect{b}=\vect{B}/B$, $v_{\parallel}=\vect{v}\cdot\vect{b}$,
$v_{\perp}=\sqrt{\|\vect{v}\|^{2}-v_{\parallel}^{2}}$, $\kappa=(\vect{b}\cdot\nabla)\vect{b}$,
and $\omega_{{\rm gyr}}$ denotes gyro-frequency. 
The second term of the right hand side denotes the curvature drift velocity and the third one the grad $B$ drift velocity. 
By using Eq. \eqref{eq:Tgyr}, 
the gyro-frequency $\omega_{{\rm gyr}}$ is given by 
\begin{equation}
	\omega_{{\rm gyr}}=2\pi/T_{{\rm gyr}}.
\end{equation}

When the pitch angle $\phi_{0}$ is small, the Larmor radius is so
small that it is quite possible to assume that the back ground magnetic
field is quasi-homogeneous. This justifies the use of Alfv\'en's guiding
center velocity. Indeed for this situations we should have $|v_{\parallel}|\gg v_{\perp}$
and the guiding center velocity can be approximated by 
\begin{equation}
\vect{v}_{{\rm gc}}\simeq v_{\parallel}\vect{b}+\frac{v_{\parallel}^{2}}{\omega_{{\rm gyr}}}\vect{b}\wedge\vect{\kappa}.\label{eq:vgc0}
\end{equation}
Let us compute the curvature drift term. Substituting $\vect{B}(r)=B_{0}\left(\vect{e}_{z}+f(r)\vect{e}_{\theta}\right)$,
$f(r)=r/q_{{\rm mag}}(r)$ into the curvature vector $\vect{\kappa}$,
we have 
\begin{equation}
\vect{\kappa}=-\frac{f(r)^{2}}{1+f(r)^{2}}\vect{e}_{r},
\end{equation}
and then, we have 
\begin{equation}
\vect{b}\wedge\vect{\kappa}=\frac{B_{0}f(r)^{2}}{1+f(r)^{2}}\left(f(r)\vect{e}_{z}-e_{\theta}\right).
\end{equation}
We therefore carry out 
\begin{equation}
\begin{split}\vect{v}_{{\rm gc}} & =v_{{\rm gc}}^{z}\vect{e}_{z}+r\Omega_{{\rm gc}}^\theta\vect{e}_{\theta},\end{split}
\end{equation}
where 
\begin{equation}
	\begin{split}	
		v_{{\rm gc}}^{z} & =\frac{v_{\parallel}}{\sqrt{1+f(r)^{2}}}\left(1+\frac{v_{\parallel}f(r)^{3}}{\omega_{{\rm gyr}}\sqrt{1+f(r)^{2}}}\right),\\
		\Omega_{{\rm gc}}^\theta& =\frac{v_{\parallel}q_{{\rm mag}}(r)^{-1}}{\sqrt{1+f(r)^{2}}}\left(1-\frac{v_{\parallel}f(r)}{\omega_{{\rm gyr}}\sqrt{1+f(r)^{2}}}\right).
	\end{split}
\end{equation}
The effective $q_{{\rm eff}}^{{\rm gc}}(r)=v_{{\rm gc}}^{z}/(R_{{\rm per}}\Omega_{{\rm gc}}^\theta)$
is therefore 
\begin{equation}
q_{{\rm eff}}^{{\rm gc}}(r)=q_{{\rm mag}}(r)\left(\frac{\sqrt{1+f(r)^{2}}+\omega_{{\rm gyr}}^{-1}v_{\parallel}f(r)^{3}}{\sqrt{1+f(r)^{2}}-\omega_{{\rm gyr}}^{-1}v_{\parallel}f(r)}\right)
\end{equation}
It should be noted that the ratio, 
\begin{equation}
\frac{q_{{\rm eff}}^{{\rm gc}}(r)}{q_{{\rm mag}}(r)}=\frac{\sqrt{1+f(r)^{2}}+\omega_{{\rm gyr}}^{-1}v_{\parallel}f(r)^{3}}{\sqrt{1+f(r)^{2}}-\omega_{{\rm gyr}}^{-1}v_{\parallel}f(r)}(>1).
\end{equation}
monotonically increases as $v_{\parallel}$ gets to be large, \textit{i.e.}
$H\simeq v_{\parallel}^{2}/2$ becomes large, for all $r$. This is
consistent with the numerical result exhibited in Fig.~\ref{fig:null}.

\begin{figure}[t]
	\centering{}
	\includegraphics[width=8cm]{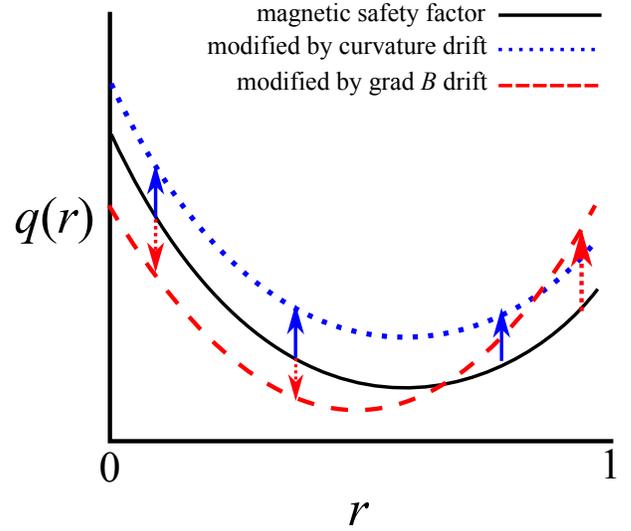} 
	\caption{(Color online) Schematic picture of $q_{{\rm mag}}$ and $q_{{\rm eff}}^{{\rm gc}}$.
	The solid curve, the dotted one, and the dashed one represent respectively
	the magnetic safety factor, the safety factor modified with the curvature
	drift effect, and the one modified with the grad $B$ drift effect.
	The blue solid and red dotted arrows denote shift brought about by curvature and
	grad $B$ drifts respectively. }
\label{fig:schematic} 
\end{figure}

We now consider the case with a large initial pitch angle, for instance $\phi_{0}=1.25$ rad. 
In this case, we need to compute the grad $B$ drift velocity, 
\begin{equation}
	\vect{v}_{{\rm gB}}=\frac{v_{\perp}^{2}}{2\omega_{{\rm gyr}}}\frac{f(r)^{2}f'(r)}{1+f(r)^{2}}\left(\vect{e}_{\theta}-f(r)\vect{e}_{z}\right).
\end{equation}
Let us introduce notations $\bar{v}_{\parallel,\perp}=v_{\parallel,\perp}/\omega_{{\rm gyr}}$
and $v_{\perp}/v_{\parallel}=\tau$. 
\begin{equation}
	\begin{split}
		v_{{\rm gc}}^{z} & =
		\frac{v_{\parallel}}{\sqrt{1+f(r)^{2}}}\left(1+\frac{\bar{v}_{\parallel}f(r)^{3}}{\sqrt{1+f(r)^{2}}}-\frac{\tau\bar{v}_{\perp}}{2}\frac{f(r)^{3}f'(r)}{1+f(r)^{2}}	\right),\\
		\Omega_{{\rm gc}}^\theta & =
		\frac{v_{\parallel}q_{{\rm mag}}(r)^{-1}}{\sqrt{1+f(r)^{2}}}\left(1-\frac{\bar{v}_{\parallel}f(r)}{\sqrt{1+f(r)^{2}}}+\frac{\tau\bar{v}_{\perp}}{2}\frac{f(r)f'(r)}{1+f(r)^{2}}\right).
	\end{split}
\end{equation}
We note $f'(r)>0$ (resp. $<0$) for $r<r_{{\rm c}}$ (resp. $r>r_{{\rm c}}$),
where $r_{{\rm c}}\simeq 0.7817$ given by solving $f'(r_{{\rm c}})=0$.
Thus, the grad $B$ drift effect decreases (resp. increases) winding
number for $r<r_{{\rm c}}$ (resp. $r>r_{{\rm c}}$). 
Since the curvature drift increases winding number, there exists $r_{{\rm c}}'>r_{{\rm c}}$
such that $q_{{\rm eff}}^{{\rm gc}}(r)>q_{{\rm mag}}(r)$ (resp. $q_{{\rm eff}}^{{\rm gc}}(r)<q_{{\rm mag}}(r)$)
for $r>r_{{\rm c}}'$ (resp. $r<r_{{\rm c}}'$). 
Let $r_{{\rm s}}$ be a critical point of $q_{{\rm mag}}(r)$, that is, $r_{{\rm s}}=1/\sqrt{2}$.
Then, $r_{{\rm c}}'>r_{{\rm c}}>r_{{\rm s}}$ and this is consistent with the numerical result exhibited in Fig.~\ref{fig:125e-2}. 
This is summarized in Fig.~\ref{fig:schematic}. 
In this case, as shown in Fig.~\ref{fig:125e-2}, the resonance shift in $r<r_{{\rm c}}'$ is larger than the shift in $r>r_{{\rm c}}'$, 
so that the resonance overlapping gets lost and the particle ITB appears for the energetic particles.

We compare $q_{{\rm eff}}^{{\rm gc}}$ with the safety factors obtained by the full orbits in Figs. \ref{fig:theory-null} and \ref{fig:theory-125e-2}.
For the small initial pitch angle cases, these are in good agreement. 
For the large initial pitch angle cases, they are good agreement qualitatively,
but not quantitatively when the particle is so energetic and is in the region $r<0.4$. 
It may be because the assumption of quasi-homogeneity is broken in for this particle. 
But the winding number $q_{{\rm eff}}^{{\rm gc}}$ and $q_{{\rm eff}}(\langle r\rangle_{\infty})$, 
and $q_{{\rm eff}}(\langle r\rangle_{T})$ are in good agreement around the resonance point on which the winding number is $2/3$, 
so that the ITB creation in chaotic field lines\cite{Ogawa2016} might be well explained by the drift effects 
explicitly appearing in the Alfv\'en's guiding center velocity \eqref{eq:vgc}. 

\section{Numerical study\label{sec:num}}

We now revisit some of the numerical observations that were done in
Ref.~\onlinecite{Ogawa2016} and that we could not explain. Using
this effective safety factor, we end up having a clear explanation
of the different phenomena. For instance when we observed that for
energetic particles with null initial pitch angle, the perturbation
effect becomes weak as energy increases, and an ITB is created, or
that the $r$-position of resonance points associated with $q=2/3$
get close to each other and they disappear, as energy increases. Another
observed and not explained feature was in the case of energetic particles
with large initial pitch angle, $\phi_{0}=1.25$ rad, we saw that
the resonances associated with $q=2/3$ shift towards the center of
the cylinder as energy increases.

\begin{figure}[tb]
	\centering{} 
	\includegraphics[width=8cm]{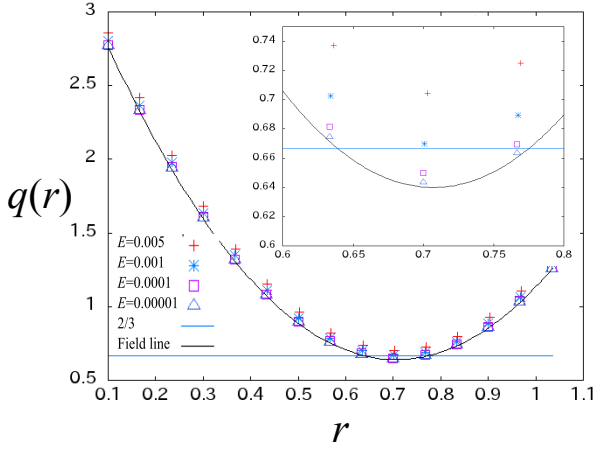} 
	\caption{
		(Color online) 
		$q_{{\rm mag}}$ and $q_{{\rm eff}}$ are plotted as functions of $r$. 
		The solid curve is $q_{{\rm mag}}$. The points give $q_{{\rm eff}}$
		for each particle with energy $E=0.00001,0.0001,0.001$, and 0.005.
		The blue straight line represents $q=2/3$, and the intersections of $q_{{\rm mag}}$
		and $q_{{\rm eff}}$ (interpolated) are resonance. 
	}
\label{fig:null} 
\end{figure}

\begin{figure}[tb]
	\centering{} 
	\includegraphics[width=8cm]{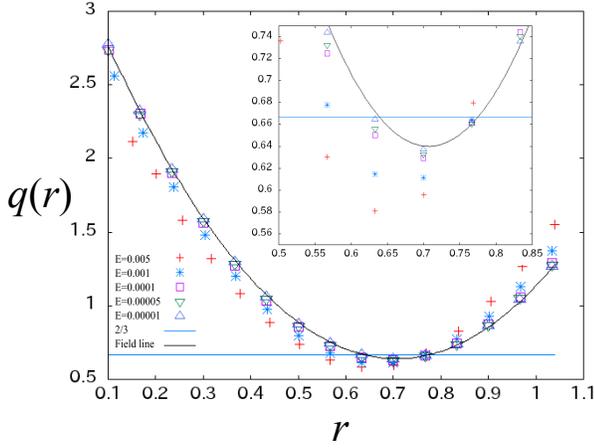} 
	\caption{ 
		(Color online) 
		$q_{{\rm mag}}$ and $q_{{\rm eff}}$ are plotted as functions of $r$. 
		The solid curve is $q_{{\rm mag}}$. The points give $q_{{\rm eff}}$
		for each particle with energy $E=0.00001,0.00005,0.0001,0.001$, and 0.005. 
		The blue straight line represents $q=2/3$, and the intersections of $q_{{\rm mag}}$ 
		and $q_{{\rm eff}}$ (interpolated) are resonance. 
	}
	\label{fig:125e-2} 
\end{figure}

Instead of taking into account the motion with the perturbation, we
shall see that these phenomena and topological changes can be explained
just using the unperturbed part, and reasoning with the effective
$q$-profile. 
One of the main advantage of this modified $q$-profile setting is that 
we can think of trajectories as if they had no Larmor radius, 
meaning as if the followed the field lines of an effective and modified magnetic field. 
In order to check this hypothesis we computed $q_{{\rm eff}}$ numerically for a relatively large finite time $T$, 
the results are displayed it in Figs.~\ref{fig:null} and \ref{fig:125e-2}. 
In Fig.~\ref{fig:null} when the pitch angle is null, 
we notice that as energy increases, the graph of $q_{{\rm eff}}(r)$ moves towards the top of the figure, 
because of this the intersections of $q_{{\rm eff}}(r)$ and the resonant condition $q=2/3$ gets to be close to each other,
and finally, disappears when $E=0.001$. This is completely in line with what is shown in Fig.~\ref{fig:particle-orbits}-(a). 
If we now look at Fig.~\ref{fig:125e-2} when the pitch angle is large ($1.25$ rad), we notice that as energy is increased, 
the graph of $q_{{\rm eff}}(r)$ moves to left, so that, the resonance points shift to left, that is,
$r$-position of the resonances decreases. 
This is also consistent with the result exhibited in Fig.~\ref{fig:particle-orbits}-(b).
We can notice as well that the resonance that is closer to the center of the cylinder shifts more than the one on the periphery, 
so that the distance between resonances gets larger. 
This is also consistent with the numerical results,\cite{Ogawa2016} the ITB gets to be wider as energy increases.

\section{Conclusion and remarks\label{sec:Conclusion}}

To conclude our study, we have shown in this paper that using an effective $q$-profile computed from trajectories 
in the unperturbed magnetic field and reasoning afterwards in the more intuitive field line setting
of a modified magnetic field was able to predict the creation or the behavior of ITBs. 
In fact, we could show that the winding number modifications induced 
by energy and pitch angle changes created a resonance shift or disappearance in the particle orbit, 
which brings about suppression of the magnetic perturbation effect in the particle motion, 
the creation of an ITB in the chaotic region of the magnetic field lines. 
These results were numerically found but left theoretically unexplained in a previous work.\cite{Ogawa2016} 
In the current paper these results are confirmed using this effective profile with first numerics based on the full particle orbits 
and as well a theory based on the guiding center velocity of Alfv\'en or using the integrable nature of trajectories in the unperturbed situations. 
We exhibit two kinds of scenarios for the creation of an ITB for the particles creation in the chaotic magnetic field lines. 
The first one is due to the resonance disappearance due to the curvature drift effect for the energetic particles with small pitch angle, 
and the second one is the resonance shift brought about by the grad $B$ drift effect for the energetic particles with large pitch angle. 
In this last case, there are two resonances for which $q_{{\rm eff}}(r)=2/3$ because of the shape of $q$-profiles. 
Although both of them shift in the same direction, resonances move towards the inside region of the cylinder, 
the distance between the two resonances gets wider, leaving enough room for an ITB to appear in the chaotic field lines.

In this paper, we have focused our study in cases when high energy brings about the creation of an ITB in a region where the magnetic ITB does not exist. 
We also would like to point out that our present results indicates that the destruction of the ITB for the energetic particles is also possible. 
Indeed let us assume that the magnetic field has an ITB between two resonances where $q_{{\rm mag}}(r)=2/3$.
When the pitch angle is small and there exists only curvature drift effect, as energy increases, 
the two resonances initially are getting closer to each other until they collide and disappear (bifurcation) then, 
it is very likely that the ITB gets destroyed in this region.
On the other hand, if when grad $B$ drift exists, unlike our case,
if the resonance shift towards the inner part of the cylinder ($r<r_{{\rm c}}'$) is smaller than the one in outer side ($r>r_{{\rm c}}'$), 
the resonances are getting also closer to each other as energy is increased and in the same spirit the ITB can be destroyed, 
with particles showing global chaos in region where the magnetic ITB is present. 

Although we use a model of magnetized plasmas with specific $q$-profile, 
the present result is applicable to more general situations. 
This is because both the particle orbit and drift velocity are determined 
by local information of the particle and field. 
Then, if the winding number approximately fitted by this kind of $q_{\rm mag}(r)$ 
we can see the same phenomenon in this area. 

We end this article by three remarks.

In the tokamak, $\theta$-dependence appears in $\vect{A}_0$, 
so that the situation is more complicated compared with cylindrical case. 
In this case, $r$-component of the velocity for unperturbed motion is not null, 
and the form of grad$B$ and curvature drift velocity are more complicated. 
Then, these drift effect for the $q$-profile may not be clarified clearly as in the cylindrical system. 
If we assume the magnetic momentum $\mu$ is the third integral of the motion, 
the particle system can be integrable when $\epsilon = 0$, \cite{Porcelli1996}
and the present result can be generalized for the tokamak.
However, it should be noted that the assumption of the third integral is sometimes broken,\cite{Cambon2014} 
but the generalized theory may work in some local region apart from the separatrix of $H_{\rm eff}$.

Secondly, it might be important to mention on the edge localized mode (ELM)\cite{Zohm1996}
which is a kind of instability appearing around the H-mode and
brings about heat and particles diffusion,
so that it breaks good confinement of plasmas in fusion reactors.
The ELM is mitigated by the resonant magnetic perturbation
called RMP which creates ergodic magnetic region with irrational $q_{\rm mag}$.
RMP is screened by the plasmas rotation with rational winding number,
and the ELM mitigation becomes not effective, so that
the precise information about resonant location of
the winding number is important to make good confinement of plasmas.\cite{Strauss2009,Becoulet2014}
As we have already seen, the particle's $q$-profile is different from
the magnetic one $q_{\rm mag}$ for various classes of particles
with energy levels and pitch angles.
Then, it might be important future work to estimate a number of particles
exhibiting resonance shift and its effect to the RMP screening.

Finally we would like to point out, that we had mentioned that the results obtained in Refs.~\onlinecite{Castillo2012, Martinell2013}
for a case with a finite Larmor radius effect in the ``trivial'' $\vect{B}=B_{0}\vect{e}_{z}$ setting but with 
an electric field with modes $\vect{E}(r,\theta,z)$ were similar to our recent result. \cite{Ogawa2016}
In fact this may be partially correct, but the main underlying mechanism is not same. 
In the situation considered in Refs.~\onlinecite{Castillo2012, Martinell2013}, 
there are neither curvature nor grad $B$ drifts, so that the effective safety factor modifications discussed in this article cannot occur,
as clearly the role of the perturbation induced by the electric field is dominant, 
while the magnetic perturbations appeared in the end as not so important in our present case. 
In this spirit, although the phenomena observed in Refs.~\onlinecite{Castillo2012, Martinell2013}
are apparently similar to the ones reported in Ref.~\onlinecite{Ogawa2016},
and since the mechanisms are different, we believe that a precise study 
when both electric field and non-uniform magnetic field can be an interesting line of research for future work.

\begin{acknowledgments}
SO thanks D. F. Escande for bringing his attention to the paper of
Fiksel \textit{et al}.\cite{Fiksel2005}
SO and XL are grateful to M. Vittot for a discussion. 
This work has been carried out within the framework of the French Research
Federation for Magnetic Fusion Studies and 
thanks to the support of the A$\ast$MIDEX project (n$^{\circ}$
ANR-11-IDEX-0001-02) funded by the ``investissements d'Avenir''
French Government program, managed by the French National Research
Agency (ANR).
\end{acknowledgments}

\end{document}